# A REVIEW OF CRITICAL INFRASTRUCTURE PROTECTION APPROACHES: IMPROVING SECURITY THROUGH RESPONSIVENESS TO THE DYNAMIC MODELLING LANDSCAPE

*Uchenna D Ani[1*], Jeremy D McK. Watson[2], Jason R C Nurse[3], Al Cook[4], Carsten Maple[5]*

[1, 2] *Department of Science Technology Engineering and Public Policy, University College London, United Kingdom,*
[3] *School of Computing, University of Kent, United Kingdom,*
[4] *Critical Insights Security Ltd, United Kingdom,*
[5] *Cybersecurity Centre, WMG, University of Warwick, United Kingdom*
* u.ani@ucl.ac.uk

## Abstract

As new technologies such as the Internet of Things (IoT) are integrated into Critical National Infrastructures (CNI), new cybersecurity threats emerge that require specific security solutions. Approaches used for analysis include the modelling and simulation of critical infrastructure systems using attributes, functionalities, operations, and behaviours to support various security analysis viewpoints, recognising and appropriately managing associated security risks. With several critical infrastructure protection approaches available, the question of how to effectively model the complex behaviour of interconnected CNI elements and to configure their protection as a system-of-systems remains a challenge. Using a systematic review approach, existing critical infrastructure protection approaches (tools and techniques) are examined to determine their suitability given trends like IoT, and effective security modelling and analysis issues. It is found that empirical-based, agent-based, system dynamics-based, and network-based modelling are more commonly applied than economic-based and equation-based techniques, and empirical-based modelling is the most widely used. The energy and transportation critical infrastructure sectors reflect the most responsive sectors, and no one Critical Infrastructure Protection (CIP) approach – tool, technique, methodology or framework – provides a *'fit-for-all'* capacity for all-round attribute modelling and simulation of security risks. Typically, deciding factors for CIP choices to adopt are often dominated by trade-offs between *'complexity of use'* and *'popularity of approach'*, as well as between *'specificity'* and *'generality'* of application in sectors. Improved security modelling is feasible via; appropriate tweaking of CIP approaches to include a wider scope of security risk management, functional responsiveness to interdependency, resilience and policy formulation requirements, and collaborative information sharing between public and private sectors.

## 1 Introduction

Critical infrastructure (CI) involves elements that are fundamental to the normal operations of the human society [1], an can be defined as referring to any asset, system or part thereof which is critical for the maintenance of vital societal functions, health, safety, security, economic or social well-being of people, and the disruption or destruction of which would have a very substantial impact as a result of the failure to maintain those functions [2]. Arguably, it may be viewed as a nation's economic "central nervous system" [3] – making it difficult for nations without a properly functional, or indeed with vulnerable CI to attain and sustain its national goals of social and economic progress and development. Examples of CIs include; Energy (electricity, oil, natural gas), Chemical, Industrial Control, Dams, Defence Industries, Emergency Services, Financial Services, Food and Agriculture, Government facilities, Commercial Services, Health and Public Health, Transportation, (Railways, Roads, Highways, Aviation, Shipping and Ports), Water and Waste water, Information Technology and Telecommunication, Nuclear [2], [4], [5].

There are growing concerns and debates about the protection of these types of CI systems, especially, how to effectively protect them given their vital positions in social and economic developments. These concerns have been highlighted with the increased emphasis on improved efficiency, performance and productivity, and this implies that CIs now rarely exist or function in isolation. Rather, they are becoming more tightly coupled into a system of (inter)dependent infrastructures, and converging with information and communications technology (ICT) and the Internet [6], [7]. This creates a complex multi-system interconnectedness and interactions referred to as a *system-of-systems (SoS)* [8].

The growing trend for convergence and multi-system interconnectedness in CIs is introducing several security issues that threaten normal economic and social functions. As new technologies such as the Internet of Things (IoT) get integrated into CNI, new security risks (threats, vulnerabilities and attacks) emerge that require specific security solutions [9]. The risks are particularly hard to identify, and handle given that the IoT has emerged from a range of disparate fields of study [10]. The benefits of CIs can be realised if they function properly and are not impaired. This requires CIs to be kept safe from harm, and secure from any disruptive or destructive compromise. Thus, it is crucial to protect CIs, especially in the light of the growing and evolving malignity. It is important to understand potential security risks and how to effectively manage them using effective protection tools and techniques.

In the above context, the objective of protection may be explored through understanding how the attributes and capabilities of existing CI protection (security) modelling approaches fit and respond to the dynamics introduced by the evolving critical infrastructure and attack ecosystems. With the increasing adoption of IoT, it is crucial to track and understand research and development directions and outcomes, together with policy and regulatory interventions, which can better support security for critical national infrastructure (CNI) systems. CNIs provide some national benefits including; supporting the attainment of a properly functioning social environment and economic markets, enhancing service security, enabling external market integrations, and allowing service recipients (consumers, clients, and users) to benefit from new and emerging technological developments [3]. As such, their safe and resilient operation is imperative and effectively protection is required. Modelling and simulation (M&S) provide a useful technique to help achieve this. In terms of CIs, M&S provide focused methods for analysing the dynamics of CI components, evaluating the interdependency and cascading effects amongst infrastructures based on system interactions [7]. M&S uses the attributes, functionalities, operations, and behaviours of CI sub-systems to support various security analysis viewpoints, recognising and appropriately managing associated security risks. With several critical infrastructure protection approaches available, the question of how to effectively model the complex behaviour of interconnected CNI elements and configure their protection as *(SoS)* remains a challenge.

This study provides novel insights on the effectiveness of existing CIP approaches (tools, techniques and methodologies) to address IoT-centric security risks, in order to guide the selection, adoption and/or development of more tailored approaches. This can provide a usable reference critical infrastructure security system developers, researchers and users. This contribution is achieved via a systematic and analytical study of available CIP tools, techniques, methodologies and frameworks herein collectively referred to as *'CIP Approaches'*. The extent of risk management coverage for CIP approaches and their security responsiveness to the dynamic modelling landscape are evaluated through investigating: (i) Common CIP modelling approaches, (ii) the industrial sectors most responsive to CIP modelling and simulations, (iii) the sub-stages of risk management mostly covered by existing CIP approaches, (iv) the extent to which resilience, (inter)dependency and policy formulation factors are considered in existing CIP Approaches.

The rest of the paper is outlined as follows: Section 2 presents the related work. Section 3 describes the methodology used in the research. Section 4 presents the results and discussion, while Section 5, concludes the article and outlines recommendations based on our research.

## 2 Related Works

There is also a growing recognition and acknowledgement that to effectively preserve operational continuity in CIs, resilience is a necessary protection objective to complement security capacities [11], [12]. Resilience can be defined as the capacity to prevent, adapt, withstand and recover swiftly from both intentional and unintentional attacks [4], [13]. There are publications [14], [15] which emphasise that the understanding, modelling, and simulation of CI attributes, functions, operations and behaviours can support security analysis, especially given dynamicity in trends and technological adoption. Prior works [16]–[19] that have explored CIP modelling and simulation approaches does not cover newer approaches, such as the Industrial Control System Cyber Defence Triage Process (ICS-CDTP) [20]. These works also fail to address emerging needs such as resilience and support for security policy updates/formulation in modern CIs. This study takes a step towards providing some answers to fill gaps in existing literature.

## 3 Methodology

To achieve our research objectives, a systematic review approach [21] was used to identify and select related and relevant literature sources. This review technique can guarantee the quality and reliability of selected articles [22].

*3.1 Literature Gathering*

Searches for relevant literature was conducted using the Web of Science (WoS) article database. WoS was chosen because of its reputation for supervised selection and inclusion of materials drawing from high-quality and high-impact indexing by humans, consistent and structured documentation, better accuracy of results, and reduced duplicates and false positives [23]. In addition, WoS is the preferred choice and standard employed by most organisations [24]. Results were restricted to peer-reviewed articles (journals, conferences, reports, books, etc.) in order to ensure quality and credibility of outcomes. Key search phrases used were: *'Critical Infrastructure Protection Tools'*, *'Critical Infrastructure Security Techniques'*, *'Critical Infrastructure Security Methodologies'*, and *'Critical Infrastructure Security Management Methods'*. Figure 1 presents the literature gathering process-flow.

*Initial search rounds* using the above terms yielded a *total search result* of 1171 articles (represented as *N*) that included duplicated articles. Exclusion filtering – represented as *e* – was done based on *titles* to identify articles related to critical infrastructure protection. Unrelated articles were discarded, and only one instance each of relevant articles was retained. 303 related articles were obtained from the process, which reflected the initial *selection study sample*; *n*.

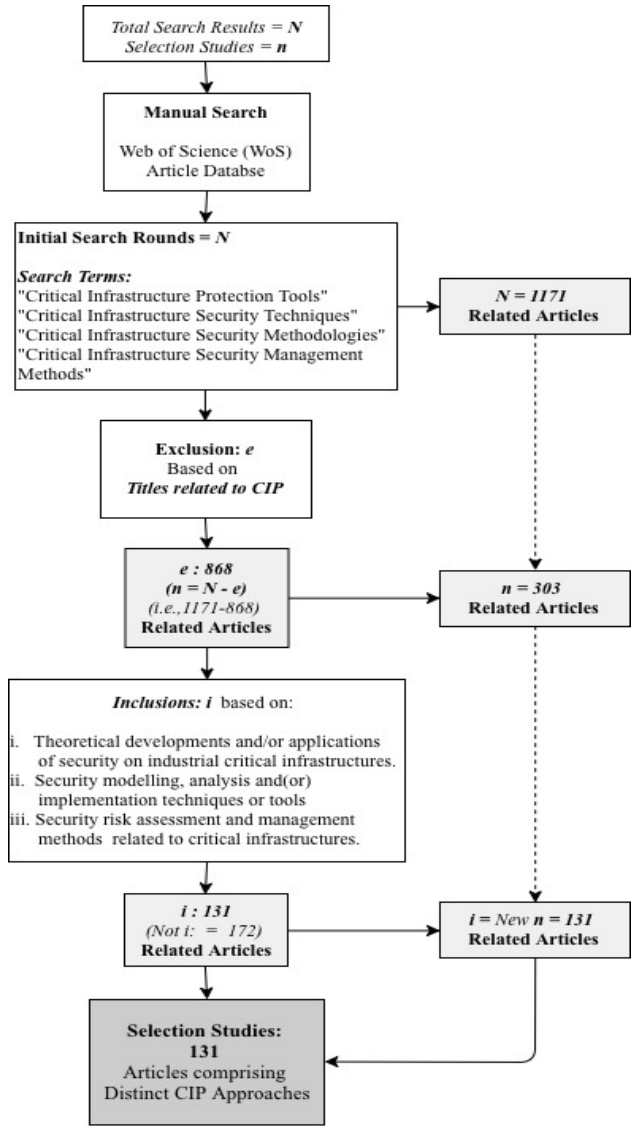

*Figure 1: Literature Gathering Process*

However, further inclusion filtering – represented as *i;* was performed to narrow the contextual scope and to select the most relevant articles that support the research objectives. These incudes:
  i.) Articles on theoretical developments and(or) applications of security on industrial critical infrastructures.
  ii.) Articles on security modelling, analysis and(or) implementation techniques or tools with use case applications in industrial critical infrastructure sectors.
  iii.) Articles on security risk assessment and management techniques/methods related to critical infrastructures.

*3.2 Literature Gathering Outcomes*
Based on the above criteria, 131 distinct CIP modelling, simulation, and/or implementation approaches characterising software tools, techniques, methodologies, and frameworks were compiled from journal and conference articles, reports, white paper, and guidelines. These are presented in Appendix A. These spanned from 1999 to 2017 and formed the final *selection study* sample (Figure 1).

*3.3 Evaluation Criteria*
The sample of the CIP approaches identified were evaluated based on criteria identified to be important for classifying CIP approaches. Most of which have been used in previous works. These criteria include; *critical infrastructure types; applicable modelling technique*; *risk management sub-stages covered;* and *(inter)dependency and resilience modelling considerations.*

Other criteria such as maturity and availability of CI tools were not used, although these have been used in the past [3] to evaluate CI tools. We think that there are uncertainties in accurately determining the maturity and availability status of some of the CIP tools given that they are mostly developed and used in-house, and as such this is an unreliable criterion to use. Reports and documentation on their use and effectiveness are not readily available in public domain. Similarly, whether they have been discarded, modified or upgraded, and at what point; is an information not easily available in the public domain. We think that adopting such criteria with potential for inaccurate data can greatly affect the accuracy of the overall study.

*3.3.1 Critical Infrastructure Type:* This is considered in order to highlight the varied levels of infrastructure criticality. In the UK, some CIs are categorised 'critical national infrastructures' – CNI, perhaps because of their huge contribution to the national economy. For example, the energy sector has an unrivalled value, a failure of which can cripple the functions of other CI sectors like emergency services, communications, health and transport, thus threatening national economy, social and political order [25].

*3.3.2 Modelling Technique:* This is considered because it connotes foundational representation of how each protection methodology is designed and applied. Although, CIP methodologies and techniques suggest varied modelling and simulation paradigms, and purpose-driven decision-making processes, they share a common goal exploring how to manage security risks. CIP modelling techniques may be classified into; *agent-based, system dynamics, empirical, network, economic*, and *other (equation-based, real-time simulation, and cellular automata) techniques* [26]. These may also be combined with additional computational methods such as discrete time-step, continuous time-step, Monte Carlo, decision-trees, geographical information systems, event monitoring, and risk management [3].

*3.3.3 Risk Management:* This context is considered because it provides a useful way of evaluating and responding to security issues in critical national infrastructure contexts [3], [16], [27], [28]. Most critical infrastructure protection implementations are typically based on risk management frameworks conceived as national or global standards [3]. Risk management approaches often vary either by; *the nature of approach*, or by *how risk is measured* [29], [30].

In this study, the *'nature of approach'* is considered (nature of approach) – emphasising the criticality of assets, the potential harm that can be done to them, and the rippling interdependency effects that can affect other connected assets on the criticality chain. Thus, supported risk management sub-stages in each CIP approach are analysed based on the National Infrastructure Protection Plan (NIPP) Critical Infrastructure Risk Management framework (RMF) [4] to underscore the purpose served by each. NIPP-RMF provides the most commonly supported guidelines in security objectives, strategies, and sector coverage. It also provides reference points to a broader community of nation states and infrastructure sectors exploring the development of tailored infrastructure security methodologies, tools and techniques [4].

NIPP-RMF suggests that risk management tools, techniques, and methodologies for CNI protection can be classified according to the purpose they serve, demonstrated by the sub-stage(s) (Figure 2) of the overall CI risk management framework that is (are) supported, while applying each approach, and the associated outputs [4]. In the framework, CI elements can be physical, cyber and human. The framework also includes a process of recurrent information sharing and feedback into subsequent risk management sub-stages. Aside from the initial sub-stage of *'setting security goals and objectives'*, the key sub-stages of the framework used as evaluation criteria are: *(i) identification of infrastructure assets*, *(ii) assessment and analysis of risks*, *(iii) risk management implementation (involving risk prioritisation, and risk control)* and *(iv) measurement of effectiveness*.

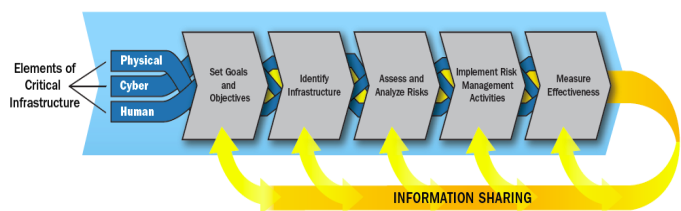

*Figure 2: Revised NIPP Critical Infrastructure Risk Management Framework [4]*

*3.3.4 Interdependency and Resilience: Interdependency* is a condition created by direct and/or indirect interconnectedness of CIs via geographically-distributed networks and physical hardware-based channels [8]. A disruption event can then spread consequences across channels of CIs, and society – technically called *'cascading effects'* [31], [32]. The harm from compromises can be physical, digital, economic, psychological, reputational, social and societal [33]–[35]. Since impact-flows across CI are as probable as the actual cyber-attacks themselves, it is crucial to understand the risks via modelling and simulations, in order to provide effective protection. Indeed, identifying and characterising interdependencies and complexities associated to CIs can improve the understanding of CIs as a SoS [36]–[38]. The insights that may be gained from failure and impact modelling and simulations can support the design of effective controls and response strategies [39].

*Resilience* describes a capacity to stop, cope with, acclimatize to, and/or recuperate from incidents that have negative consequences [4], [12]. With resilience in CIs, infrastructure functions, operations and services are reasonably maintained even in the face of an infrastructure disruption or compromise [33]. With attacks that cause cascades and failures, impacts can be economically and socially massive, so the need to be well-equipped to withstand and recuperate from adverse events is ever more necessary. Quite often, CI incidents happen unexpectedly, and complete control is rarely feasible. The dynamic threats and hazards landscape is such that it is hardly possible to foresee, prevent, prepare for, or control all CI security incidents, which in most cases can be unknown or emergent [33]. This necessitates a shift from the usual crisis management to resilience management to address supply chain disruptions and rapid restoration.

Appropriate readiness and recovery requires strengthening and investing in resilience to minimise sub-system vulnerabilities to restrict occurrence, intensity and propagation of failures and impacts on CIs and in turn on society [12]. It is appropriate that 'resilience' is becoming fundamental in general crisis and disaster management discourse, and is the focus of widespread efforts for resisting, absorbing, accommodating and recovering from the effects of security threats. This emphasises effort on preventive, mitigative and preparedness activities prior to a CI security crisis, response during the crisis, and the recovery after the crisis [13]. Integral dependencies and failure cascades should be considered in analysing and designing for resilience, and they should underscore the whole cycle of a CI security crisis, since it is impractical to guard against all threats [40], [41].

## 4 Results and Discussions

CIP approaches were analysed based on obtained information obtained about them from bibliographic literatures: reports, articles, white papers and guideline to arrive at informed insights. The list of approaches in Appendix A reflects a wide range of research being conducted in the area of critical infrastructure protection and considered relevant in the light of keeping pace with new trends like the IoT. The results of evaluating 131 CIP modelling approaches are thus presented.

*4.1 Results*

*4.1.1 Common modelling techniques applied in the CIP approaches*

For modelling techniques, we find that a variety of techniques are in use as shown in Figure 3. Empirical-based modelling appears to attract the widest application with 36 (27.3%) of reviewed approaches in its favour. Examples of approaches in this category include: *HURT, FTA, RVA,* and *RMCIS.* Network-based modelling is seen in 32 (24.2%) approaches including *CASCADE, IRRIIS,* and *HAZOP*. System dynamics-based modelling is used in 20 (about 15.2%) approaches including *AIMSUN, CIPMA,* and *ICS-CDTP*. Agent-based modelling is applied in 26 (19.7%) approaches including, *CIMS, ADVISE, N-ABLE* and *GoRAF*. Some CIP approaches combine two or more techniques. For example, *ACT* combines economy-based and system dynamics-based modelling, while *ADVISE* and *GoRAF* both combine agent-based and real-time simulation techniques.

*4.1.2 Industrial sectors most responsive to CIP modelling and simulations*
Results of sector-based classification shown in Figure 4 indicates that 72 (54.5%) CIP approaches applied to *energy* comprising *electricity, pipeline & oil, natural gas sectors*. 42 (37.1%) CIP approaches applied to the *transportation* sector. *Water & Waste Water* has 47 (35.6%) and *Chemical* sector has 41 (31.1%) among others. *Emergency Services* has the least with 13 (9.8%) applicable CIP approaches.

From a multi-sector applications viewpoint, 111 (84.1%) of the CIP approaches mainly cover up to 5 sectors (1-5 sectors). 60 out of the 111 provide software support for engaging and implementing their designed operational processes. Only 4 (3%) of the CIP approaches mainly cover between 6 to 10 sectors. These include: *IIM, Risk Map, RVA,* and *BLDMP*. Of the CIP approaches. 16 (12.1%), cover at least 11 CI sectors including; *Athena, BIRR, CASCADE, CIDA, CIMSuite, CIPDSS-DM, EURACOM, Fort Future, IRRIS, ACT* and *ICS-CDTP*.

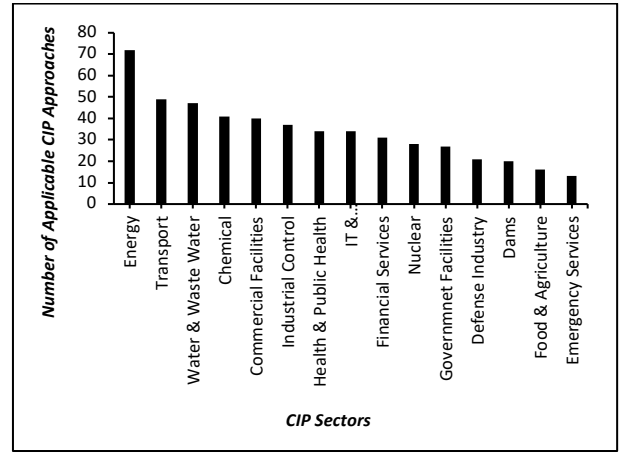

*Figure 4: Sector-based Analysis of Occurrence of Critical Infrastructure Modelling and Protection Approaches*

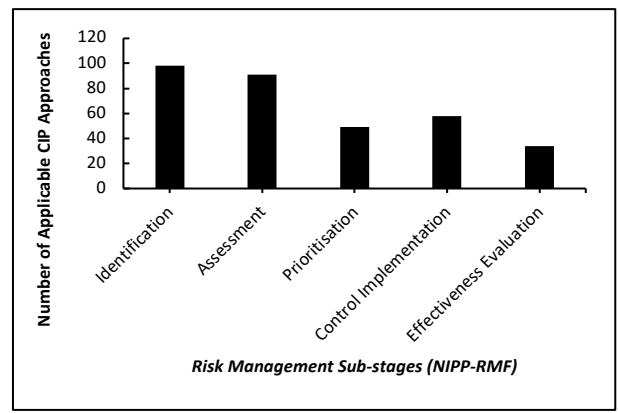

*Figure 5: Risk Management Stages Covered*

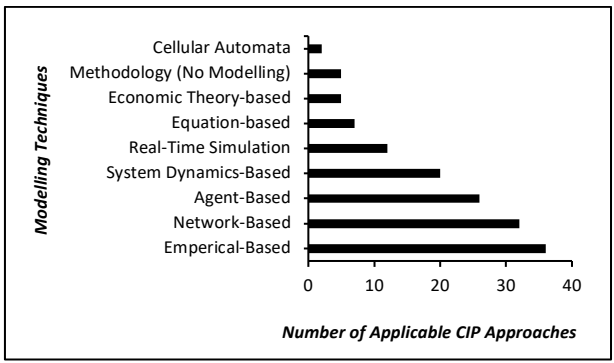

*Figure 3: Modelling Techniques*

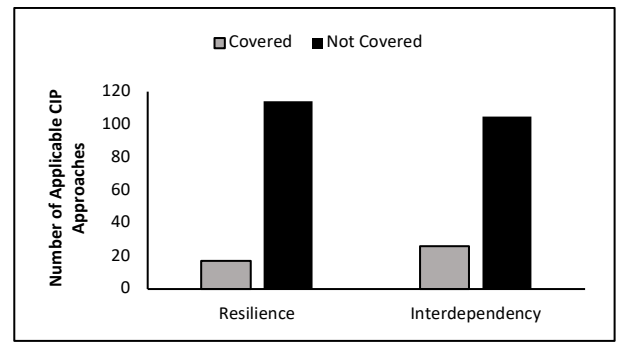

*Figure 6: Analysis of Interdependency and Resilience Characteristics*

*4.1.3 Risk management sub-stages covered*
98 (74%) CIP approaches (Figure 5) included aspects related to *'identification of critical infrastructures and vulnerabilities'* sub-stage. 91 (68%) CIP approaches considered the *'assessment of risks'* sub-stage. 82 (62%) CIP approach considered some sort of *'risk management implementation'*, which is a sub-stage comprising *'risk prioritisation'* and *'risk control'* based on the *Revised NIPP Critical Infrastructure Risk Management framework* [4]. From these, 20 (nearly 24%) approaches considered *'risk prioritisation'* alone without *'control implementations'*, while 29 (nearly 35%) approaches considered *'risk control implementation'* alone without any form of prioritisation. 34 (nearly 26%) CIP approaches considered *'effectiveness evaluation'* sub-stage of the revised NIPP-RMF.

Looking at the risk management implementation sub-stages (*prioritisation* and *control*) in isolation, more CIP approaches (58) i.e., 44% of 131 characterised a sort of *risk control implementation* than *risk prioritisation* which had 49 CIP approaches. The 58 tools also represent 54% of the CIP approaches categorised under the risk implementation stage. Only half (29) approaches considered both risk prioritisation and control implementation stages, the remaining half considered just one of the two stages. In addition, fewer approaches consider all five sub-stages exclusively. 49% of the CIP approaches considered only two sub-stages, which most typically include: *risk identification and risk assessment*. Only 6 (4.5%) of CIP approaches considered all five sub-stages. These include: *BIRR, COUNTERACT, EURACOM, IRRIIS, NSRAM*, and *NIPP-RMF* itself. This latter category of approaches exists as broad guiding frameworks (*NIPP-RMF*) or methodologies (*BIRR, COUNTERACT, EURACOM, IRRIIS*) or a complex modelling tool with software support (*NSRAM*). The methodologies and framework are mostly applicable to a generality of CI sectors, while the software modelling tool NSRAM is applicable for chemical, energy and IT/Communications sectors.

*4.1.4 Interdependency and Resilience considerations*
Only 28, or 21%, of reviewed approaches considered elements of *interdependency modelling* (Figure 6). Examples of approaches with explicit characterisation of interdependencies include: *AIMS, Athena, CASCADE, CIPMA, CISIA, N-ABLE, NEMO, and UIS*. We assume these considerations of interdependency phenomena may be connected to the core objectives for developing the tools in the first place. This may be influenced by an acknowledgement of criticality of interdependencies for both constructive and destructive impacts on CI operations. Behavioural and cascading effects of functions and failures appear to be at the core of the objectives for developing most of the listed approaches in this group.

For *resilience coverage*, only 18 (14%) out of the 131 CIP approaches clearly considered aspects related to resilience modelling (Figure 6). Examples in this category include: *BIRR, CIMS, CIPMA, DECRIS, EURACOM, Fort Future, IIM, MBRA, HAZOP, Risk Maps, RAMCAP-Plus, Sandia Risk Assessment Methodology, NIPP-RMF, and RMCIS*. Potential reasons why a greater proportion of CIP approaches do not consider resilience attributes may relate to: (i) the core objectives of their developments, which points to perceived requirements for protecting critical infrastructures. *Resilience may not have been part of the requirements for development.* (ii) the development time for the CIP approaches, which may have predated the concept of resilience in critical infrastructures – *CI resilience may not have been clearly defined and/or gained wide attention.* Thus, it is assumed that the CIP approaches that have considered resilience may have emerged as responses to newer challenges associated with convergence and hyper-connectivity trends, which have made resilience a necessary objective. Only 4 (3.1%) CIP approaches considered both resilience and (inter)dependency modelling attributes, as well as policy and regulatory formulation attributes.

*4.2 Discussion of Findings*
On a general base, a huge proportion of the sample of CIP approaches exist as either tools, techniques or methodological frameworks. The approaches seem mostly structured to handle operations and performance modelling and simulations, rather than security. However, they can also be used for security-related attributes such as evaluating the impact of security features or their absence with critical infrastructures. In addition, it is not clear how much of IoT performance and security are reflected in the reviewed CIP approaches. This may be because most of the approaches predate the IoT trend. A common characteristic shared by these the various CIP approaches is that they are all based on risk management, although they are distinguishable by the scope or sub-stage of overall security risk management functions considered in each approach.

We find that the ***common modelling techniques which receive widespread interest and adoption include: agent-based, system dynamics-based, network-based and empirical-based***. These are not the only applicable techniques but only represent the most commonly used, and a quite applicable for IoT contexts too. Newer techniques may be defined from combining two or more of the above, or even entirely new modelling paradigms depending on intended development goals. The aim being to enable a multi-level modelling and simulation to meet the various requirements that may be set by IoT.

At the moment, *the empirical-based model seems to be most widely employed for CIP research and the development of security approaches*. This is closely followed by network-based and system dynamics-based modelling techniques. These preferences could link to growing data-driven technology trends and reflect an increasing need to understand and evaluate CI security using real or actual historic data and drawing from expert knowledge and experiences.

For IoT-based critical infrastructures, this is good news, because IoT systems typically characterise huge volume of data from sensors connected to CI components. Empirical modelling can enable a simulation system involving the collection and dissemination of such sensor-acquired CI data and their management via a context-aware data distribution service to be used by application [42]. Potential applications include smart cloud services that can take and integrate such data with other available data (crowd-sourced or crowd-sensed) to support real and high-fidelity analysis and insights. Although this will typically come with a lot of complexities – networks, communications and pervasiveness that needing resolve during the modelling process. By combining sensed and historic data, it becomes easier to observe and connect the interrelationships and interdependencies among critical infrastructure components, system and sectors. Thus, empirical-based modelling techniques seem to better support the identification of more recurrent, realistic and suggestive failure patterns, quantifying interdependency-related indicators for risk mitigations, supporting emergency decision-making and providing validation parameters to support other modelling techniques [26].

Agent-based modelling is also very applicable for IoT as 'things' in critical infrastructures can be considered as agents that interact with other agents to enable a seamless operation. Network-based modelling provides an ability to model interdependencies among IoT-based CI systems, especially within localised areas. The technique simplifies the derivation of insights related to CI representations along topological or flow-pattern analysis line, and the evaluation of cascading impacts. As the emphasis on fidelity, dependency and resilience of CIs increases, interests and support may move towards empirical-based modelling, with progressive support for network-based, agent-based, and system dynamics-based modelling. The lesser emphasis on other modelling techniques (such as equation-based, economy-based, etc.) may indicate a lack of popularity for their concepts perhaps due to a greater complexity in their use.

Although modelling and simulation are applicable to several industrial CI sectors, *the widest or highest sensitivities seem to come from the energy sector.* This sector comprises *electricity, pipeline and oil, and natural gas CIs. Transportation, water & waste water, and chemical industries also have significant interest and responsiveness.* These sectors all fall within the category of CNIs defined in the UK CPNI documentation [5]. The energy sector is critical because several other critical infrastructures depend on its products/services. Nearly all other CIs require some form of energy source (from an energy sub-sector) to drive their functionalities. Thus, the energy sector takes on an almost indispensable position within critical infrastructure interdependencies. The consequences and impact of energy infrastructure failure can inevitably ripple through and affect other dependent infrastructures. If unattended to, the effects can cause a myriad of damaging cascading outcomes *(physically, operationally,* and *economically)* in the chain of interdependent CIs within a national or global social ecosystem. This can explain the greater concerns and sensitivities concerning a more secure and continuously improving energy sector, as opposed to the lesser emphasis in other CIs.

Although the IoT promises huge benefits for CIs like energy and transport, it also brings new and complex issues. IoT poised to improve energy and transport efficiencies, reliability, proactive maintenance, and utilisation visibilies amongst other. But there are issues of security, interoperability, scalability and logistics needing to be address. Modelling provides ways of learning the scales on both perspectives, and how they might be addressed.

*Emergency services, food &. agriculture, dams appear to demonstrate low interests and responsiveness to CIP modelling and simulations.* This could be because these sectors often appear at the tail of CI interdependency chains, and typically have low direct and immediate large-scale social consequences and impacts when compromised. In addition, responses in form of solutions are typically responsive to, and trail similar paths as the direction of growing malicious events. Moreover, these sectors appear not to be suffering increasing cyber-attacks judging by known recorded incidents [43]. While it may be rational to focus solutions where there are greater threats and risk challenges, the potentials for common cause failures are very imminent, hence, it is crucial to also give a good measure of attention other CI sectors.

Clearly, ***empirical-based, network-based, and agent-based modelling techniques are the three most widely used techniques in fulfilling the risk management framework sub-stages.*** *System-dynamics-based techniques* are typically used *for simulating continuous system behaviours* such as *estimating the effectiveness of implemented procedures in critical infrastructures*. Observed consistent patterns suggest that *empirical-based modelling is more widely employed in risk identification, assessment, prioritisation* and *control implementations sub-stages*, and *less consideration for effectiveness evaluations sub-stage*. *Network-based modelling* is widely adopted for risk *identification, assessment* and *control implementation*. We assume this is so because network-based techniques (either topology-based or flow-based methods) are helpful in capturing interdependency characteristics, CIs descriptions, and identifying the critical components with suggestions for emergency protection and response improvements [26]. As IoT continue to find its way into CIs, concerns about how to learn their impact on host components and interdependencies are likely to influence the use of network-based modelling techniques.

In general, results suggest *a greater interest on CI researches around risk identification, assessment, and control implementation in that order*. Traits of *risk prioritisation* and *effectiveness evaluation do not seem to enjoy much attention and research*. Thus, the aspects of security risk management with the most interests and presumed relevance, as well as where the direction of CIP developments tends can be easily seen. It reveals that beside identifying and assessing security risks on CIs, the next thought in the minds of CI owners and operators is *what controls to implement to mitigate or eliminate characterised risks*. This indicates that thinking about implementing controls may be viewed as more important than first understanding the varied criticality levels of security risks before conceiving a strategy for implementing controls to yield the highest possible security and resilience outcomes. These thoughts need to start shifting towards the new risk status that relate to IoT-in-CIs; from identification to effectiveness evaluations. New thought directions for modern CIs (now with IoT) risk management need to extend to the less appreciated significance of assessing the extent to which desired CIP has been achieved post control implementation to obtain success measures and potential guides for further improvements. Common issues and debate seems to pivot around the variability of approaches for measuring risks, and how such measures are reflective of true system states also need to address IoT issues.

Quite ***a few (less than one-fifth) of the CIP approaches considered resilience***. *Greater focus and purpose-driven developments tended towards defending malicious attacks or compromises on CIs.* Contexts that can help reduce attack impacts and sustain operations or functionalities during and after successful compromises did not enjoy wide consideration. This could be because *most CIP approaches were evolved before the trend of 'resilience' – implying they were already in use prior to when resilience became a feature of significance and concern, and newer/updated versions of these approaches are yet unavailable*. Alternatively, the significance of ensuring resilience may also not have been clearly understood by developers as at the time of developing some of the approaches - making for why resilience feature is not reflected in the tools. Besides implementing preventive security, the need to ensure a capacity to continue operations – delivering the needed services while managing attacks – is ever more important. Clearly, for CNI sectors, disruptions or their cascading effects are not welcomed phenomena. *Anticipation, prevention, absorption, adaptation* and *rapid recovery* are essential towards achieving resilience. While newer CIP approaches benefit from resilience features, older approaches need to be modified to embody resilience where missing. As an example, we consider the NIPP-RMF that lacked resilience in its maiden version but was later revised in 2013 to include both security risk and resilience characteristics.

Although not having widespread acceptance and inclusion, ***(inter)dependency coverage seems more prevalent than resilience in the CIP approaches.*** The modelling dimensions covered include: *Component/infrastructure-level* (e.g., AIMS, MUNICIPAL), *operational/functional-level* (e.g., CASCADE, CIPMA), vulnerability-level (e.g., MIA), *Cost/Time-dependencies* (e.g., CI3), and *market effects (e.g., CommAspen)*. Results reveal a growing appreciation of the significance of interdependencies in CIP modelling and analysis. Private and public CI owners and operators are beginning to recognise that learning the relationships amongst CI components and systems can greatly support the attainment and enhancement of security and resilience.

Again, the few approaches mostly emerge from research institutes and academic institutions rather than government regulatory agencies. Current outlook also suggests that *the domain of CIP is more characterised with self-garnered defensive solutions rather than being compliance-based mediums*. This individuality and the differences in security problems and requirements can be the drivers for the development and adoption of bespoke protection techniques by infrastructure organisations.

From a multi-attribute consideration viewpoint, ***very few CIP approaches currently consider the combined attributes of interdependency and resilience.*** For example, out of 131 approaches reviewed, *only CIMS, CIPMA, and IIM appear to satisfy the above criteria*. Interestingly, none of these approaches applied the more common empirical or network-based modelling techniques. Rather, these use agent-based and system dynamics modelling. This reveals the limitations of existing CIP approaches to sufficiently address the security dynamics in modern CIs. The level of multi-attribute coverage appears to be significantly low compared to the proportion of CIP approaches being developed, further suggesting a crucial need to upgrade or refine existing approaches to address any attribute deficiencies in order to improve their effectiveness.

## 5   Conclusion and Recommendations

Arguably, modelling for critical infrastructure protection seems not entirely new, as its underlying concepts typically relate to safety modelling and analysis. Only that over time, has security become relevant and emphasised due to technology trends. This has made CI sectors readily susceptible to intentional cyber-engineered attacks. Having also become so tightly coupled and interdependent, incidents show that the compromise, disruption and failure of CIs is not only restricted to causes and vectors related to natural disasters. Human-initiated actions via technology abuse or mal-interventions can be and increasingly are, an influence.

However, what seems new and perhaps not well reflected – at least directly in most of the critical infrastructure modelling and security approaches (tools, techniques, and methodologies) – is the concept of addressing *'resilience'*. Most CIP approaches reviewed mainly focus on exploring concepts and phenomena related to *security, reliability, dependability* and *risks* in CIs. We reckon that a plausible reason for this may be linked to the early and more widespread emphasis on these attributes. Also, it may be attributed to *the ease in defining and evaluating the above attributes compared to evaluating resilience*. For example, studies [44] indicate a common acknowledgement by power company executives about a better comparative convenience for the ease of defining and measuring of CI reliability than CI resilience. Be this as it may, this apparently intractable attribute is now hihgly relevant to meeting the evolving protection needs of CI sectors.

The typical contexts that characterise the objectives for developing CIP modelling approaches – either tools, techniques or methodologies; emphasise the desire to understand the dynamic behaviours of CI systems using modelling techniques such as *agent-based, system dynamics-based, network-based,* and *empirical-based techniques*. These techniques help to identify and characterise the causes of functional/operational anomalies and/or disruptions within CI setups through determining critical hazards and risks, their interdependencies, consequences and impact cascades. CIP should embrace modelling and analysis of security-related operations, activities and risk management, mostly within the confines of specific infrastructure environments and sectors.

In relations to how CIP is modelled, security risk management methods drive the process for gaining deeper security-related performance insights for CIs to support effective responses. Wider interests focus more on the starting sub-stages of risk management including: *(i) identification of Critical infrastructure, hazards and vulnerabilities*, and *(ii) assessment and analysis, of security risks*. Empirical-based modelling combined with risk identification, assessment, implementation and management of risk are among the most common implementation modes. These seem influenced by the growing adoption of setups and models that generate or feed-on actual scenario data to support CIP sensitivity analysis for decision-making.

The energy and transportation sectors demonstrate the widest concerns and efforts on protecting CNIs. This is not surprising as these sectors have higher criticalities and provide services that sustain several other sectors. Less sensitive sectors like emergency services, food and agriculture and dams need to emulate the drives and actions of the energy and transportation sectors, to ensure that they are well-equipped to handle security threats when they eventually surface. In balancing the trade-off between *'specificity'* and *'generality'*, their key individual benefits need to be considered. Specificity allows for more focused context coverage and analysis, which can mean better and more tailored solutions, which will mean better and more tailored solutions. Conversely, generality enables a tool to be applicable to multiple CI sectors. However, a single approach – a tool or technique – cannot support holistic security modelling of CIs. A combination of multiple approaches – preferably integrated into a tool and technique (methodology or framework) – is perhaps the way to go. In addition, an approach that includes the pragmatics of implementing necessary control actions to curb security risks, as well as learning the effectiveness of controls can be a preferred solution.

Resilience modelling links to interdependency, and interdependency analysis contributes information and insights about the degree of cross-systems impact inducible by failures or disruptions. It also contributes to the perception on the level of resilience achievable in principle and practice. While some of the CIP approaches acknowledge and consider dependency or cross-dependency relationships and attributes, a larger proportion either implicitly include it or utterly overlook it. In this era of IoT, advancing technology convergence and system hyper-connectivity, understanding the interdependencies amongst CI components and systems can strongly make the difference between *ignorance* and *knowledge* on the nature, type, and degree of resilience required to enhance protection of CIs.

A significant number of the CIP approaches reviewed emerge as instruments developed by government agencies responsible for protecting CIs, or by research laboratories, for example the Idaho National Laboratory in the USA (who are also funded by the government. This could present an effective way for government and regulators to encourage a wider use of CIP approaches. However, the extent to which government-sponsored approaches are realistically adopted in the private sectors is often unclear or in doubt, since this latter group run substantial chunks of CI systems. We posit that policies that support easy adoption of approaches are necessary in the above regard. Also, the private sector shows mixed responses to seamless monitoring and reporting of cybersecurity vulnerabilities and incidents. Even in cases where shared information, cooperation and assistance can significantly support realising acceptable protection of CIs from threats and attacks, decisions and actions are often determined by growing pressures for business competitive advantage [45]. Policies that advance and manage shared and collaborative information capability between public (government) and private sector stakeholders would be very helpful in this regard.

Clearly, security risks for CIs are evolving along new technology pathways where IoT devices and applications are finding their ways into CI systems. IoT systems are commonly characterised by; *variability of scale in components*, *temporality of connections amongst devices*, and *the heterogeneity of actors*. These characteristics influence conditions that can exist between periodic risk assessments without necessarily reporting on instantaneous risk impacts to the whole system. Since the risk assessment mode in existing CIP approaches are designed to operate statically periodically, they lack appropriate capabilities to address the dynamics and transient threats in IoT [30], [46], agree that dynamic risk assessment has become necessary. Such dynamic assessment mode *would need to address this looming problem by catering for emerging system connectivity in real-time, as well as characterising, in clear and timely way, the level of temporality of devices in relations to their risk impacts*.

Thus, to ensure effective; CIP modelling and simulations, sensitivity to dynamic trends, and potentially sustainable and efficient 'Living in the IoT', this study recommends that,

i. Other security conscious but less-responsive critical infrastructure sectors such as *emergency services, food & agriculture, and dams;* should draw lessons from the efforts of the energy and transportation sectors. Analogous approaches should increase the ability to evaluate and understand security risks to attendant infrastructures and operations. They can support better understanding of any associated dependencies and cascading impacts and improve understanding of how to establish effective security and resilience. The decision-making processes related to measuring the effectiveness of readiness activities and investments will be improved, as well as the behavioural responses to CI disturbances or disruptions in the sectors.

ii. Newer or updated CIP modelling approaches should *be developed or revised to capture scope of IoT in security risk management – from identification to effectiveness evaluations.* This is to support appropriate *alignments and responsiveness to the evolving trends* introduced by *new technologies such as IoT and IIoT*. Such approaches also need to *adopt dynamic and real-time assessment processes* to address the issues introduced by IoT in CIs, and the high impact security risks that emerge.

iii. A *strong public-private sector partnership is important and should be vigorously pursued by both stakeholder groups to achieve better security and resilience in CIs*. Such collaboration can empower the public sector to monitor, in timely and efficiently ways, and to aggregate information about CI security threats, vulnerabilities, incidents and impacts as they emerge. The public sector can also provide the risk information to private sector operators to help them ensure an informed and well-organised security management.

## 5 Acknowledgements

The Research leading to the results presented in this paper comes from the Analytical Lenses for Internet of Things (ALIoTT) project under the PETRAS Cybersecurity of the Internet of Things Research Hub. It has received funding from The Engineering and Physical Sciences Research Council (EPSRC).

Appendix A: Selection Study Sample of Critical Infrastructure Protection Approaches

| CIP Approaches | Full Meaning | Purpose Description | Web Link |
|---|---|---|---|
| ACT | Attack Countermeasure Tree | Tool for developing attack scenarios, identification and selection of best countermeasures. | *https://ieeexplore.ieee.org/document/5466633/* |
| ActivitySim | Activity Simulator | Used for modelling the activity representation of US population | *http://permalink.lanl.gov/object/tr?what=info:lanl-repo/lareport/LA-UR-08-07134* |
| ADVISE | Adversary-Driven State-based System Security Evaluation | Tool for simulating attacks on systems, and evaluating the probability of attack success. | *https://www.perform.illinois.edu/Papers/USAN_papers/10VAN02.pdf* |
| AIMS | Agent-based Infrastructure Modelling and Simulation) | Used for analysing the behaviour of interdependent critical infrastructure systems | *http://ebagheri.athabascau.ca/papers/ijbpim.pdf* |
| AIMSUN | Advanced Interactive Microscopic Simulator for Ur-ban and Non-Urban Networks | Used for Traffic Modelling and Simulation | *https://www.aimsun.com/aimsun-next/* |
| AMTI Loki Toolkit | Advanced Modelling & Techniques Investigation (Loki Toolkit) | Used for modelling and studies of complex adaptive system of systems related to critical infrastructure interdependencies | *http://prod.sandia.gov/techlib/access-control.cgi/2012/121117.pdf* |
| AT/FP | Anti-Terrorism/Force Protection | Modelling and planning the perimeter and waterway security of ships in ports | *https://savage.nps.edu/RobotTelemetry/DonCioXmlWgNpsSlides/NPSATFPProjectFlyer.2007Apr19.pdf* |
| ATAV-SCADA | Attack Trees for Accessing Vulnerabilities in SCADA (Canada) | Tool for calculating the characteristics of the highest attack event | *https://www.semanticscholar.org/paper/The-Use-of-Attack-Trees-in-Assessing-in-SCADA-Byres-Franz/02fa72c0bfd76c731201156f81c40952b9da80d1* |
| Athena | - | Used for modelling, identifying and ranking most dependent components/nodes, component/infrastructure vulnerability analysis, direct, cumulative and cascading impacts of changes to infrastructure systems. It also identifies cascading, cumulative, direct and indirect effects on nodes. Used for developing dependency and consequence reasoning support to the critical infrastructure (transportation) architecture. | *https://ieeexplore.ieee.org/stamp/stamp.jsp?tp=&arnumber=5067457* |
| ATOM | Air Transportation Optimization Model | Used for modelling and evaluating the consequences of partial or total outage at an airport or set of airports for a prolonged period of time. | *https://books.google.co.uk/books?id=YtXvAgAAQBAJ&pg=PA32&lpg=PA32&dq=Air+Transport+Optimisation+Model+-+ATOM&source=bl&ots=JGVn-y2lvK&sig=3dEXuQBYKh-FrstfbM_wvgvNJ_4&hl=en&sa=X&ved=2ahUKEwio57WJnabcAhWQxIUKHZdCkoQ6AEwB3oECAIQAQ#v=onepage&q=Air%20Transport%20Optimisation%20Model%20-%20ATOM&f=false* |
| BIRR | Better Infrastructure Risk and Resilience | Used for assessing vulnerabilities and reporting of risks | *http://www.dis.anl.gov/projects/ri.html* |
| BLDMP | Boolean Logic Driven Markov Processes | Tool for modelling attacks, characterizing and quantifying potential sequences and steps for attacks. | *https://www.sciencedirect.com/science/article/pii/S0951832017301850* |
| BMI | Protection of Critical Infrastructures – Baseline Protection Concept (German Government) | A Methodical plan for risk identification, assessment and control in critical infrastructure domains through cooperation between public and private infrastructure operators. | *https://www.preventionweb.net/files/9266_2967ProtectionofCriticalInfrastruct.pdf* |
| CAPRA | Comprehensive Approach for Probabilistic Risk Assessment | Used for modelling, assessing and reporting disaster risk from a probabilistic point of view | *https://www.ecapra.org* |
| CARVER2 | Criticality Accessibility Recoverability Vulnerability Espyability Redundancy | Used for modelling and prioritization of threats and terrorist targets | *http://publications.jrc.ec.europa.eu/repository/bitstream/JRC70046/lbna25286enn.pdf* |
| CASCADE | | Used for modelling and analysis of cascading disruptions and failures in large and interconnected infrastructures | *https://ieeexplore.ieee.org/stamp/stamp.jsp?tp=&arnumber=1385362* |
| CEEESA | Centre for energy, environmental, and economic systems analysis (Argonne National Laboratories) | Tools for analyzing network vulnerabilities, modelling gas flows and infrastructure losses | *https://ceeesa.es.anl.gov* |
| CERT Initiatives | EU CERT group members: | Methodologies for adopting and implementing security teams and capabilities for managing and protecting national critical infrastructures | *http://www.egc-group.org/contact.html* |
| CERT/CSIRT | Computer (emergency) security incident response team (Carnegie Mellon University) | Tool for monitoring, identification, and prevention of computer security and related incidents | *https://resources.sei.cmu.edu/asset_files/Handbook/2003_002_001_14102.pdf* |
| CI3 | Critical Infrastructure Interdependencies Integrator | Used for modelling and estimating the time and costs for partial or complete restoration of critical infrastructures after disruptions or failures | *http://www.ipd.anl.gov/anlpubs/2002/03/42598.pdf* |
| CIDA | Critical Infrastructure Dependency Analysis tool | Used for modelling and analysis of the dynamics of cascading failures with time. Also used to model and analyze interdependencies and risk reductions | *https://github.com/geostergiop/CIDA/wiki* |
| CIMS | Critical Infrastructure Modeling System | Analysis of risk and visualization of cascading impacts of operational anomalies. Used for sensitivity analysis, policy, regulations, and response planning. | *http://www.dis.anl.gov/projects/ri.html* |
| CIMSuite | Critical Infrastructure Modelling Suite | Used for proactive modelling of critical infrastructure targeted disruptions (natural and human-initiated). | *http://www4vip.inl.gov/factsheets/docs/cimsuite.pdf* |
| CIP/DSS | Critical Infrastructure Protection Decision Support System | Used for comparative modelling and analysis of risk mitigation strategies on individual infrastructures. Uses scenario-based impact analysis results. | *http://public.lanl.gov/dp/CIP.html* |
| CIPDSS-DM | Critical Infrastructure Protection Decision Support System Decision Model | Used for modelling decision-making under risks and uncertainty conditions | *http://www.ipd.anl.gov/anlpubs/2008/12/63060.pdf* |
| CIPMA | Critical Infra-structure Protection Modeling and Analysis | Used for evaluating failures, dependencies and resilience of critical infrastructure, as well as cascading impacts on other infrastructures. Supports the development of policies and regulations for national security | *http://www.dis.anl.gov/projects/ri.html* |
| CISIA | Critical Infrastructure Simulation by Interdependent Agents | Used for modelling agents/system interdependencies, and analysis of emergency responses and their origin. | *http://www.chiarafoglietta.com/wp-content/uploads/2015/04/Cisia.pdf* |

| CIP Approaches | Full Meaning | Purpose Description | Web Link |
|---|---|---|---|
| COMM-ASPEN | Agent-based simulation model of the US economy | Used for modelling the effects of market decision and disruptions of telecommunications infrastructure to the economy. | *http://www.dis.anl.gov/projects/ri.html* |
| CORAS-BRA-SCADA | CORAS-Based Risk Assessment for SCADA (USA). | Tool for modelling the risks of ICS prototypes using the CORAS framework | *https://pdfs.semanticscholar.org/3143/940955a76a49646ba2954e0735a0ec18d7ca.pdf* |
| COUNTERACT | Cluster of User Networks in Transport and Energy relating to Anti-terrorist Activities | Used for risk assessment, mitigation and reporting | *http://www.dis.anl.gov/projects/ri.html* |
| CSASG-SCADA Systems with Game Models | Cyber Security Analysis of Smart Grid SCADA Information Security (USA) | Tool for identifying the best action strategy for attackers and defenders, and relative payoffs. | *https://dl.acm.org/citation.cfm?id=2602089* |
| CSRA-NPP | Cyber Security Risk Assessment in Nuclear Power Plants (Korea) | Tool for identifying and characterizing risk assessment activities at initial design stages | *http://koreascience.or.kr/article/ArticleFullRecord.jsp?cn=OJRHBJ_2012_v44n8_919* |
| Cy-T SCADA RF | Cyber-Terrorism SCADA Risk Framework (Australia) | Measuring cyber-terrorism threats and implementing control measures | *http://ro.ecu.edu.au/cgi/viewcontent.cgi?article=1004&context=isw* |
| DECRIS | Risk and Decision Systems for Critical Infrastructures | Used for risk and vulnerability analyses that focus on critical infrastructure (drinking water, energy supply, transportation, ICT) interdependencies. | *https://www.sintef.no/projectweb/samrisk/decris/* |
| DEW | Distributed engineering workstation (Electrical Distribution Design, Inc. Sponsored by DOE and DoD) | Tool for identification and analysis of interdependencies, asset management, and operations planning for power systems. | *https://www.eee.hku.hk/~cees/software/dew.htm* |
| DMRIM-SCADA System | Digraph Model for Risk Identification and Management in SCADA System (USA). | Tool for vulnerability identification, faults and failure diagnosis, and risk impact assessment. | *https://ieeexplore.ieee.org/document/5983990/* |
| DUTCH NRA | Dutch Government | Tool used for analyzing threats and hazards using multi-criteria decision making techniques to achieve reduction of risks. | *https://english.nctv.nl/binaries/poster-st-geneva-2015-analyst-network-(8)_tcm32-84227.pdf* |
| EAR-PILAR | National Cryptology Centre Spain | A tool for asset characterization, risk (threats, vulnerabilities, and impacts) modelling, and control evaluations. Considers identification, classification, ratings, and dependencies amongst assets | *http://www.pilar-tools.com/en/tools/pilar/v71/index.html* |
| ECI-GIS | Geographic information systems and risk assessment (EU sponsored Joint Research Centre). | A tool for modelling operational continuity following loss and damage of critical infrastructures. | *https://core.ac.uk/download/pdf/38613171.pdf* |
| EMCAS | Electricity Market Complex Adaptive System | Used for modelling and evaluating operational and economic impacts of various external events on complex power systems (e.g. electricity) | *https://www.energyplan.eu/othertools/national/emcas/* |
| EpiSimS | Epidemic Simulations | Used for modeling and analysis of the spread of diseases | *http://public.lanl.gov/sdelvall/p556-mniszewski.pdf* |
| EPRAM | Electric Restoration Analysis Tools | Used for modelling electric power restoration | *http://www.mssanz.org.au/modsim2013/D2/stamber.pdf* |
| ERC-SCADA System-Petri Net Analysis | Evaluating the Risk of Cyber Attacks on SCADA Systems via Petri Net Analysis. | Tool for evaluating operational risks using non-probabilistic metrics approach. | *https://ieeexplore.ieee.org/document/5168093/* |
| EURACOM | European Risk Assessment and Contingency Planning Methodologies for Interconnected Energy Networks | All-hazard risk assessment and contingency scheduling. | *http://www.dis.anl.gov/projects/ri.html* |
| FAIT | Fast Analysis Infrastructure Tool (Sandia National Lab, sponsored by US DHS) | Knowledge base tool (including emergency network and georeferencing data) for performing economic impact analysis across multiple critical infrastructure sectors. | *http://www.dis.anl.gov/projects/ri.html* |
| FastTrans | Los Alamos National Lab | A parallel microsimulator tool for transportation networks for simulating and routing very large numbers of vehicles on real-world road networks in a fraction of real time. | *https://www.lanl.gov/orgs/adtsc/publications/science_highlights_2011/docs/6InfoSciPDFs/sunil.pdf* |
| FEPVA | Framework for Electricity Production Vulnerability Assessment (Los Alamos National Lab) | Tool for assessing the potential impact of natural disasters or malicious attacks for both response and preventative purposes. Specifically used to determine the power plants with impact potentials and the extent feasible. | *https://www.gpo.gov/fdsys/pkg/GOVPUB-C13-f3de19ca7b535ba3207a5be512241f84/pdf/GOVPUB-C13-f3de19ca7b535ba3207a5be512241f84.pdf* |
| FINSIM | Financial System Infrastructure (Los Alamos National Lab) | Tool for modelling financial service sector as a complex decentralized system with multiple interacting autonomous decision nodes or agents such as banks, traders, markets, and brokers. | *https://cnls.lanl.gov/annual26/abstracts.html* |
| FMEA/FMECA | Failure Mode Effect and Criticality analysis | Technique for analyzing probable system failures, enumerating potential impacts, and classifying control and mitigation actions. | *https://pdfs.semanticscholar.org/aba3/1bf32898f29ea56be2e1f5b4f99938face35.pdf* |
| Fort Future | US Army Corps of Engineers | A tool that follows a multiple simulation approach for multi-criteria decision support. Used for simulating test plans for Department of Defense installations, and evaluating a set of alternatives. | *https://ascelibrary.org/doi/pdf/10.1061/40794%28179%2922* |
| FTA | Fault Tree Analysis | A deductive technique for evaluating risk causes from a combination of inputs. | *http://asq.org/quality-progress/2002/03/problem-solving/what-is-a-fault-tree-analysis.html* |
| GAMS-CERO ERA | Enterprise Risk Assessment | Technique for managing and mitigating risk using administrative procedures and resources. | *http://www.dis.anl.gov/projects/ri.html* |
| GIS Interoperability | Geographical Information Systems Interoperability | A methodology for emergency coordination and support using geographical information systems. | *https://books.google.co.uk/books?id=eoB6nTkhLgkC&pg=PA388&lpg=PA388&dq=Challenges+for+the+application+of+GIS+interoperability+in+emergency+management&source=bl&ots=A9AYBmqk0n&sig=EaYUOn_X24FOaIYX3rXlAvFVyuw&hl=en&sa=X&ved=2ahUKEwjRqIHz67XdAhUHLewKHTCiBo0Q6AEwAHoECAAQAQ#v=onepage&q=Challenges%20for%20the%20application%20of%20GIS%20interoperability%20in%20emergency%20management&f=false* |
| GoRAF | University of New Brunswick (Canada) | A tool for critical infrastructure resource identification, and metric-based estimation of economic losses. | *https://www.inderscienceonline.com/doi/pdf/10.1504/IJRAM.2007.015297* |
| HAZOP | Hazard and Operability Analysis | Technique for system examination and risk management based on theory of assumptions that risk events occur due to deviations from design and operating plans. | *http://pqri.org/wp-content/uploads/2015/08/pdf/HAZOP_Training_Guide.pdf* |
| HCSim | Healthcare Simulation (Los Alamos National Lab) | A modelling tool for assessing the impact of mass casualties in health care and public health institutions (e.g., hospitals) | *https://permalink.lanl.gov/object/tr?what=info:lanl-repo/lareport/LA-UR-13-24605* |

| CIP Approaches | Full Meaning | Purpose Description | Web Link |
|---|---|---|---|
| HM-BRMCI | Hierarchical, model-based risk Management of Critical Infrastructures | Tool for automating the definition of risk mitigation plans and activities. | https://www.sciencedirect.com/science/article/pii/S0951832009000349 |
| HURT | Hurricane Re-location Tool (Los Alamos National Lab) | A tool for modelling the relocation of Hurricane | http://www.lanl.gov |
| HYDRA Pop & Eco Modeling | (Los Alamos National Lab) | Integrated service-oriented architevture tool for modeling and simulating infrastructures with seamless interoperability. | https://public.lanl.gov/rbent/hydra-with-cover.pdf |
| I2SIM | Infrastructures Interdependencies Simulation (University of British Columbia) | A tool for simulating scenarios for disaster responses at system level with impact characterization. | http://www.ece.ubc.ca/%7Ejiirp/ |
| ICS-CDTP | Industrial Control System Cyber Defense Triage Process | Tool for threat analysis, attack modelling, and control and countermeasure applications | https://www.sciencedirect.com/science/article/pii/S0167404817301505 |
| IEISS | Interdependent Environment for Infra-structure System Simulations (University of Virginia) | A modelling tool for simulating electricity and natural gas flow, outage characteristics, and system interdependencies. | http://www.bwbush.io/projects/ieiss.html |
| IIM | Inoperability In-put-Output Model (Sandia National Labs and Los Alamos National Labs) | A tool for sector-based economic impact analysis of infrastructure attacks and failures. | https://ascelibrary.org/doi/pdf/10.1061/%28ASCE%291076-0342%282005%2911%3A2%2867%29 |
| Infrastructure Disruptions | - | Tool for modelling the state of infrastructure systems under abnormal conditions, and evaluating the economic consequences of abnormalities. | http://www.dis.anl.gov/projects/ri.html |
| INTEPOINT VU | IntePoint LLC | A modelling tool that combines various techniques for complex environments analysis and system-wide interdependencies modelling across physical, virtual and social networks. | https://www.nist.gov/sites/default/files/documents/el/msid/Critical_Infrastructure.pdf |
| IRAM | Infrastructure risk analysis model (US Military Academy) | Tool used to model and simulate resource allocation for interconnected infrastructure reliability. Used for risk quantification. | https://ascelibrary.org/doi/10.1061/%28ASCE%291076-0342%282000%296%3A3%28114%29 |
| IRAM-SCADA INFORMATION Sec | Improved Risk Assessment Method for SCADA Information Security (Serbia) | Evaluating the effectiveness of intrusion, detection, and prevention systems in controlling attacks. | http://eejournal.ktu.lt/index.php/elt/article/view/8027/4033 |
| IRRIIS | Integrated Risk Reduction of In-formation-based Infra-structure Systems (IRRIIS Project, EU) | Interdependency and resilience modelling, analysis and management of critical infrastructures | https://www.irriis.org |
| Knowledge Mgt & Visualisation | Carnegie Mellon University | Tool for analyzing vulnerabilities related to the distribution of fuel | https://inldigitallibrary.inl.gov/sti/3489532.pdf |
| LogiSims | Los Alamos National Laboratory | Tool for modelling and planning preparation for a disasters and concurrent responses to a disaster | http://public.lanl.gov/rbent/bent-pes.pdf |
| LS-DYNA | Livermore Software Technology Corporation | A tool for modelling large complex system structures and behaviours related to failures such as: changing boundary conditions, deformations, crashes and explosions. | http://www.lstc.com/products/ls-dyna |
| LUND | University of Lund (Sweden). Sponsored by the International Energy Agency | Grounded Network theory methodology for modelling the relationships between nodes in a system of roads or rail interconnected transport infrastructure. | https://www.iea.lth.se/publications/Theses/LTH-IEA-1061.pdf |
| MARGERIT V2 | Spanish Ministry for Public Administrations | methodology for Risk Analysis and Management for security of computer systems, digital and data networks. | https://www.enisa.europa.eu/topics/threat-risk-management/risk-management/current-risk/risk-management-inventory/rm-ra-methods/m_magerit.html |
| MBRA | Model-Based Risk Assessment (Naval Postgraduate School, Center for Homeland Defense & Security) | Analysis of critical infrastructure network components and faults for efficient resource allocation | https://www.chds.us/ed/items/2164 |
| MIA | Methodology for Interdependency Assessment | A methodology for identifying and characterizing critical interdependencies of the systems in relations security vulnerabilities. | https://link.springer.com/content/pdf/10.1007%2F978-3-642-21694-7_1.pdf |
| MIITS | Multi-Scale Integrated Information & Telecommunications System (Los Alamos National Laboratories) | A tool for simulating high fidelity network topology, internet communication sessions and packets, and actual scalability representations. | https://ieeexplore.ieee.org/document/4117861/ |
| MIN | Multi-layer Infrastructure Networks (Purdue University) | A simulation tool for solving flow equilibrium and optimal budget allocation problem related to automobile, urban freight and data network layer | https://link.springer.com/content/pdf/10.1007%2Fs11067-005-2627-0.pdf |
| Modular Dynamic Model | Sandia National Laboratory | A tool for modelling and simulating energy infrastructure interdependency operations including generation, transmission, distributions and trading. | https://www.sandia.gov/nisac-ssl/wp/wp-content/uploads/downloads/2012/04/a-modular-dynamic-simulation-model.pdf |
| MSM | MIT Screening Methodology (MIT = Massachusetts Institute of Technology) | A methodology for prioritizing vulnerabilities | http://www.dis.anl.gov/projects/ri.html |
| MUNICIPAL | Multi-Network Interdependent Critical Infrastructure Program for Analysis of Lifelines (Rensselaer Poly-technic Institute, USA) | A decision support tool simulating infrastructure moving parts and interdependencies within coastal regions to define optimal response before, during and after hazards. | http://eaton.math.rpi.edu/faculty/Mitchell/papers/decisiontechnologies.pdf |
| N-ABLE | National A-gent-Based Laboratory for Economics (Sandia National Laboratories and Los Alamos National Laboratories) | A tool for analyzing economic factors, responses and downstream consequences of infrastructure interdependencies | http://www.dis.anl.gov/projects/ri.html |
| NEMO | Net-Centric Effects-based Operations Model (Sparta, Inc.) | A tool for modelling impact cascades of events through multiple infrastructure networks, and determining the results of course of actions. | http://www.dodccrp.org/events/10th_ICCRTS/CD/papers/128.pdf |

| CIP Approaches | Full Meaning | Purpose Description | Web Link |
|---|---|---|---|
| Neptune Tides | Neptune Navigation Software (UK) | A tool for simulating wind speed and analysis of flood surges. | *http://www.neptunenavigation.co.uk/tides.htm* |
| Net-Centric GIS | York University | A tool that used to support decision making propositions using GIS interoperability features. | *https://inldigitallibrary.inl.gov/sti/3489532.pdf* |
| NEXUS Fusion Framework | IntePoint, LLC | A tool for modelling and visualizing planned and unplanned effects and consequences of an event through multiple infrastructures | *https://www.oii.ox.ac.uk/research/projects/nexus/* |
| NGAT | Natural Gas Analysis Tools (Argonne National Laboratories) | A tool for modelling natural gas pipeline infrastructures | *https://inldigitallibrary.inl.gov/sti/3489532.pdf* |
| NGFast | Natural Gas Fast (Argonne National Laboratory) | A tool for simulating natural gas systems, and impact assessment of pipeline breaks or failures. | *https://ieeexplore.ieee.org/stamp/stamp.jsp?arnumber=4419711* |
| NIPP-RMF | National Infrastructure Protection Plan – Risk Management Framework (Dept. of Homeland Security) | A process methodology for risk management for protecting critical infrastructures. It combines threats, vulnerability and consequence analysis to drive prioritization of effective controls to minimize impacts. | *https://www.dhs.gov/xlibrary/assets/NIPP_RiskMgmt.pdf* |
| NSRAM | Network Security Risk Assessment Method (James Madison University) | Analysis of cyber and physical infrastructure security risks, determining the response nature of system to attacks and incidents | *https://works.bepress.com/george_h_baker/12/download/* |
| NSRM | Network Security Risk Model | Methodology used to support the selection of risk management countermeasures and controls | *https://onlinelibrary.wiley.com/doi/epdf/10.1111/j.1539-6924.2008.01151.x* |
| OGC CIPI | Critical Infrastructure Protection Initiative (Open Geospatial Consortium) | A methodology for managing emergency incidents through inter-agency data exchange and alert notifications | *http://www.opengeospatial.org/projects/initiatives/cipi1.2* |
| PCI-Information | Projects of Common Information (Joint Research Centre, Sponsored by the European Commission) | A methodology for standardizing energy communication systems of European Union stakeholders and regulators | *https://ec.europa.eu/energy/en/topics/infrastructure/projects-common-interest* |
| PFNAM | Petroleum Fuels Network Analysis Model (Argonne National Laboratories) | A tool for hydraulic computation of crude oil and petroleum products transportation via pipelines. | *http://www.gss.anl.gov/publications-2/* |
| PipelineNet | US Federal Emergency Management Agency and the Environmental Protection Agency | A GIS-based tool for modelling the flow and concentration of contaminants in water pipeline infrastructures. Also used to estimate risks to public water supply. | *https://www.tswg.gov/sites/default/files/publications/PipelineNet%20TB.pdf* |
| PMU-Based RAFPCS | PMU-Based Risk Assessment Framework for Power Control Systems (USA) | Tool for real-time monitoring cyber intrusion impacts on the behaviours/dynamics of power systems. | *https://ieeexplore.ieee.org/stamp/stamp.jsp?arnumber=6672731* |
| QCRREM | Quantitative Cyber Risk Reduction Estimation Methodology (USA) | Tool for evaluating risk reductions in an enhanced security SCADA System. | *https://ieeexplore.ieee.org/document/1579754/* |
| QCSRAM-SCADA | Quantitative Cyber Security Risk Assessment Methodology | Tool for assessing vulnerabilities from historic data related to threats, asset value, and outage costs | *https://www.scientific.net/AMR.960-961.1602* |
| QMACSR-SCADA Systems | Quantitative Methodology to Assess Cyber Security Risk of SCADA Systems (Korea) | Tool for calculating cyber threats expected damage | *https://www.scientific.net/AMR.960-961.1602* |
| QTRIM | Quantitative Threat-Risk Index Model (Idaho National Engineering and Environmental Laboratory) | Tool used for evaluating security risks in relations to terrorist attacks against national infrastructures. | *https://inldigitallibrary.inl.gov/sites/sti/sti/2535260.pdf* |
| QualNet | Scalable Network Technologies, Inc | A tool for modelling and analysing the behaviour of real communications networks. | *https://web.scalable-networks.com/qualnet-network-simulator-software* |
| R-NAS | Railroad Net-work Analysis System (Sandia National Laboratories and Los Alamos National Laboratories) | A tool for modelling the impacts to the flow of commodities over the rail network and infrastructure in the US, especially when one or more components of the rail system are unavailable. | *https://www.sandia.gov/nisac-ssl/wp/wp-content/uploads/RNAS-20160119_SAND2016-1408M.pdf* |
| RA-SCADA Railways | Risk Assessment in GPS-based SCADA for Railways (USA) | Identification of the origin of risks | *https://ac.els-cdn.com/S0167404815001388/1-s2.0-S0167404815001388-main.pdf?_tid=dd19f4ca-8664-4262-86fb-3a4c728f32a1&acdnat=1534163806_56cd404652f32b69a028bf6a20a63b3d* |
| RADR | Risk Assessment Detection and Response | Identifying sensors with high priorities for prioritizing security budgets | *https://pdfs.semanticscholar.org/ee2e/e3dca15c4836b07c7a0e2c265329a9298901.pdf* |
| RAIM | Real-time Monitoring, Anomaly detections, Impact analysis, and Mitigation Strategies SCADA security framework | A tool for real-time monitoring and anomaly detection, impact analysis and security control implementations in power control SCADA infrastructure networks. | *https://ieeexplore.ieee.org/document/5477189/* |
| RAMCA | Risk-Assessment Model for Cyber Attacks. | Tool for calculating summed losses on revenue related to cyber-attacks. | *https://pdfs.semanticscholar.org/8a41/a48819b6ecf62424bb4d6041a8a31a630cfe.pdf* |
| RAMCAP-Plus | Risk Analysis and Management for Critical Asset Protection Plus (American Society of Mechanical Engineers) | A methodology for the assessment of risk and resilience and prioritization across all critical infrastructure sectors | *http://files.asme.org/ASMEITI/RAMCAP/17978.pdf* |
| Restore | Interdependent Repair and Restoration Processes (Argonne National Laboratories) | A tool for modelling the restoration and recovery of critical infrastructure systems from incidents. Used to estimate time and cost attributes of restoration goals. | *http://www.anl.gov/egs/group/resilient-infrastructure/resilient-infrastructure-capabilities* |
| Risk Maps | Risk Mapping, Planning and Assessment. | A Methodology for systematic risk inventory management including support planning to reduce risk impacts | *https://www.fema.gov/risk-mapping-assessment-and-planning-risk-map* |
| RMGCIS | Risk Management Guide for Critical Infrastructure Sectors (Canada) | A methodology for risk and resilience assessment and control implementations. | *https://www.publicsafety.gc.ca/cnt/rsrcs/pblctns/rsk-mngmnt-gd/rsk-mngmnt-gd-eng.pdf* |
| RTDS | Real Time Digital Simulator (RTDS Technologies Inc) | A tool for real-time simulating and testing the changing behavior of power systems. | *https://www.rtds.com* |

| CIP Approaches | Full Meaning | Purpose Description | Web Link |
|---|---|---|---|
| RVA | Risk and Vulnerability Analysis (Danish Emergency Management Agency) | A methodology for analyzing threats, vulnerabilities and risks in critical infrastructure sectors. It also supports prioritization for effective vulnerability and risk controls. | *http://brs.dk/eng/inspection/contingency_planning/Documents/RVA-model_user_%20guide.pdf* |
| S-RAM | Risk Assessment methodology (Sandia National laboratory) | A methodology for automated assessment of risks and resilience related to physical critical infrastructure attacks | *https://prod.sandia.gov/techlib-noauth/access-control.cgi/2008/088143.pdf* |
| SAIV | Security of Activities of Vital Importance (French Government) | Methodology for protection critical infrastructures based on private-public sector discussions, and priority-based support of security across critical infrastructure sectors. | *https://www.ssi.gouv.fr/en/cybersecurity-in-france/ciip-in-france/faq/* |
| SC-Based ARAC | Scenario-based Approach to Risk Analysis in Support of Cyber Security (USA) | Used to support effective resource allocation in finances and personnel for critical attacks | *https://inis.iaea.org/search/searchsinglerecord.aspx?recordsFor=SingleRecord&RN=43118741* |
| SessionSim | Los Alamos National Laboratories | A tool for generating realistic communication sessions or data traffic | *https://ieeexplore.ieee.org/document/5429274/* |
| SIERRA | System for Import/Export Routing and Recovery Analysis (Sandia National Laboratories and Los Alamos National Laboratories) | A tool for modelling and estimating flow diversion between ports. | *https://www.osti.gov/servlets/purl/1142053* |
| SRM-ICSP | Security Risk Methodology for Instrumentation and Control System Processes | Tool for assessing cyber risks for nuclear instrumentation and control systems using Bayesian networks and event tree modelling techniques. | *https://www.sciencedirect.com/science/article/pii/S1738573316302935* |
| TEVA | Threat Ensemble Vulnerability Assessment (EPA) | A tool for analysing the vulnerabilities of water distribution systems, measuring public health and economic impacts, and modelling threat mitigation and response strategies. | *https://ascelibrary.org/doi/abs/10.1061/40737%282004%29482* |
| TIMQAV-CIS | Two Indices Method for Quantitative Assessment of the Vulnerability of Critical Information Systems | Tool use to support informed decisions about countermeasures related to security vulnerabilities. | *https://www.sciencedirect.com/science/article/pii/S0268401208000054?via%3Dihub* |
| TRAGIS | Transportation Routing Analysis Geographic Information System (Oak Ridge National Laboratories) | A tool for modelling transportation (rail, waterway and highway) routing | *https://web.ornl.gov/sci/gist/TRAGIS_2005.pdf* |
| TranSims | Transportation Analysis Simulation System (Los Alamos National Laboratories) | A tool for simulating vehicular movements, and analyzing the consequences of urban transportation system. | *https://code.google.com/archive/p/transims/* |
| UIS | Urban Infrastructure Suite (Los Alamos National Lab) | A tool for simulating interactive urban infrastructures, their behaviours and effects of interdependencies. | *http://www.sandia.gov/nisac/uis.html* |
| UML-CI | University of New Brunswick, Fredericton, Canada | A reference method for modelling infrastructure systems high-level metamodels to aid system profiling and management. | *https://link.springer.com/article/10.1007/s10796-008-9127-y* |
| UPMoST | Urban Population Mobility Simulation Technologies (National Infrastructure Simulation and Analysis Center) | A tool used to model the movement of entities across multiple domains and interfaces. | *https://ieeexplore.ieee.org/abstract/document/1265180/* |
| USArmy Risk Mitigation | Los Alamos National Laboratory | A tool for simulating the management of fresh water network infrastructure in relations to usage at U.S. military bases. | *https://www.systemdynamics.org/assets/conferences/2001/papers/Lee_MA_1.pdf* |
| VACSPI | Vulnerability Assessment of Cyber Security in Power Industry | For estimating cyber vulnerability indices of infrastructures in the power sector. | *https://ieeexplore.ieee.org/document/4076075/* |
| VAM-SCADA Security | Vulnerability Assessment Methodology for SCADA Security | Tool for assessing vulnerabilities and the security of SCADA system | *https://inldigitallibrary.inl.gov/sites/sti/sti/3562811.pdf* |
| VINCI | Virtual Interacting network Community (University of Pisa, Italy) | A tool for modelling secure network management architecture for critical infrastructures using virtualization capabilities. | *https://ieeexplore.ieee.org/document/5628730/* |
| VISAC | Visual Interactive Site Analysis Code (Oak Ridge National Laboratory) | A tool for analysing accidents/incidents at nuclear or industrial facilities, and modelling the range of damaged and downtime. | *https://www.visac.ornl.gov/HelpFiles/iitsec02.html* |
| WISE | Water Infrastructure Simulation Environment (Los Alamos National Laboratories) | A tool for infrastructure and interdependency analysis of water and waste water flows. | *https://ascelibrary.org/doi/10.1061/40792%28173%2958* |